\documentclass[10pt, a4paper]{article}
\usepackage[final]{lrec2026}

\usepackage[T1]{fontenc}
\usepackage[utf8]{inputenc}
\usepackage{graphicx}
\usepackage{booktabs}
\usepackage{tabularx}
\usepackage{amsmath}
\usepackage{amssymb}
\usepackage{multirow}
\usepackage{subcaption}
\usepackage{siunitx}
\usepackage{pifont}
\usepackage{makecell}
\usepackage{adjustbox}
\usepackage{tcolorbox}
\usepackage{xcolor}

\PassOptionsToPackage{hyphens}{url}
\usepackage{hyperref}


\newif\ifshowchanges
\showchangestrue    
\showchangesfalse   

\newcommand{\chg}[1]{%
  \ifshowchanges
    \textcolor{red}{#1}%
  \else
    #1%
  \fi
}

\def\acknowledgmentsname{Acknowledgments}

\ProvideDocumentEnvironment { acknowledgments } { +b } {%
  \section*{\acknowledgmentsname} { #1 }
} {}

\newcommand{\pval}[1]{%
    \num{#1}%
    \ifdim #1pt < 0.001pt ***%
    \else\ifdim #1pt < 0.01pt **%
    \else\ifdim #1pt < 0.05pt *%
    \fi\fi\fi%
}

\title{Human-Centered Multimodal Fusion for Sexism Detection in Memes with Eye-Tracking, Heart Rate, and EEG Signals}

\name{Iván Arcos\textsuperscript{1}, Paolo Rosso\textsuperscript{1,2}, Elena Gomis-Vicent\textsuperscript{1}}

\address{
\textsuperscript{1}PRHLT Research Center, Universitat Politècnica de València (UPV), 46022 Valencia, Spain\\
\texttt{iarcgab@etsinf.upv.es, prosso@dsic.upv.es, egomvic@upvnet.upv.es}\\
\textsuperscript{2}ValgrAI – Valencian Graduate School and Research Network of Artificial Intelligence, 46022 Valencia, Spain
}

\abstract{
The automated detection of sexism in memes is a notoriously challenging task due to multimodal ambiguity, cultural nuance, and the use of humor to provide plausible deniability. As a result, content-only models often fail to capture the complexity of human perception. To address this fundamental limitation, we introduce and validate a human-centered paradigm that augments standard content features with rich physiological data. We created a novel resource by recording Eye-Tracking (ET), Heart Rate (HR), and Electroencephalography (EEG) from 16 subjects (8 per experiment) while they viewed \chg{3984} memes from the EXIST 2025 dataset. Our statistical analysis reveals significant physiological differences in how \chg{subjects} process sexist versus non-sexist content. Sexist memes were associated with higher cognitive load (evidenced by increased fixation counts and longer reaction times), and with differences in EEG spectral power across the Alpha, Beta, and Gamma frequency bands. This pattern, commonly linked in previous research to increased attentional engagement and cognitive effort during visual processing, suggests that sexist memes may elicit more demanding neural activity compared to non-sexist ones.
Building on these findings, we propose a novel multimodal fusion model that integrates these physiological signals with enriched textual-visual features derived from a Vision-Language Model (VLM). Our final model achieves an AUC of 0.794 in binary sexism detection, a statistically significant 3.4\% improvement over a powerful VLM-based baseline. The fusion of physiological data proves particularly effective for nuanced and ambiguous cases, boosting the F1-score for the most challenging fine-grained category, \textit{Misogyny and Non-Sexual Violence}, by an unprecedented 26.3\%. Our work demonstrates that human physiological responses provide a robust, objective signal of perception that can significantly enhance the accuracy and human-awareness of automated systems for countering online sexism. \\
\newline \Keywords{sexism detection, memes, physiological signals, eye-tracking, electroencephalography, heart rate, multimodal fusion, vision–language and physiological resources}
}

\begin{document}

\maketitleabstract

\section{Introduction}

The proliferation of social media has reconfigured communication, but it has also created a fertile environment for the dissemination of harmful content. Online sexism and misogyny have become pervasive phenomena that normalize discriminatory attitudes, silence the voices of women, and cause tangible psychological harm \cite{unwomen_repository, geh_ucsd}. The scale of the problem is staggering: a 2025 Amnesty UK report revealed that 73\% of Gen Z users have witnessed misogynistic content online, with TikTok being the primary vector (70\% of respondents) \cite{amnesty_uk_toxic_tech}. This abuse is not trivial and leads to lowered self-esteem and self-censorship \cite{nomore_org}, constituting a significant human rights issue in the digital age.

Within this ecosystem, memes have emerged as a key medium of communication, but also as a vehicle for propagating sexist ideologies \cite{issac_karmaveer_sexism_memes}. The challenge for their automatic detection lies in their unique semiotic structure: their meaning emerges from the multimodal interplay between text and image and is often shrouded in humor, irony, or satire \cite{reunir_sexist_humour}. This strategic ambiguity creates "plausible deniability," allowing creators and sharers of harmful content to evade accountability by claiming it was "just a joke." This feature makes sexist memes a particularly difficult problem for artificial intelligence systems \cite{ut_cs_female_astronaut_misogyny}.

Current computational models, even state-of-the-art ones, systematically fail when faced with this challenge. The "Hateful Memes Challenge" by Facebook AI demonstrated that cutting-edge multimodal models performed little better than random chance when classifying memes where the offensive nature arose from the combination of individually non-offensive text and images \cite{kiela2020hateful}. This failure highlights a fundamental gap: "sexism" is not an intrinsic property of the content but an emergent phenomenon shaped by human interpretation. Human perceptual and affective responses provide complementary information that can help models capture the nuanced and context-dependent nature of sexist content more effectively \cite{arxiv_explaining_sexism_detection}.

This work proposes a human-centered pivot: instead of \chg{training an AI model} to replicate subjective labels, we propose teaching it to recognize unconscious physiological responses that manifest in a human who perceives a given content as sexist. Physiological and neural reactions—such as gaze patterns \cite{acl_anthology_gaze4hate}, heart rate variability \cite{plosone_rhrveasy}, or brain activity \cite{researchgate_physiological_signals_affective_computing}—are objectively measurable and less ambiguous than conscious interpretations. This study leverages physiological data from eye-tracking (ET), heart rate (HR), and electroencephalography (EEG) as a more reliable proxy for the human experience, addressing the following research questions:

\begin{enumerate}
  \item[\textbf{RQ1}] \textit{Are physiological responses—eye-tracking, heart rate, and electroencephalography—significantly different between sexist and non-sexist memes, and across distinct types of sexism?}

  \item[\textbf{RQ2}] \textit{Are there statistically significant differences in performance among unimodal classifiers based solely on (a) textual, (b) visual, and (c) physiological-derived features when detecting and categorizing sexist memes?}

  \item[\textbf{RQ3}] \textit{Can the incorporation of physiological data into a multimodal fusion model (text + image) yield a significant improvement in classification performance over a baseline text\chg{+}image model?}
\end{enumerate}

\paragraph{Contributions}
\begin{enumerate}
    \item \textbf{A new resource.} A multimodal dataset built from \chg{3984} memes with rich physiological responses from 16 subjects (8 per experiment) across two large-scale studies: 
\textbf{Experiment 1} recorded \textbf{ET and HR} responses (\textbf{7782} recordings), while 
\textbf{Experiment 2} recorded \textbf{EEG and HR} responses (\textbf{7714} recordings), totaling \textbf{15496} instances.

   \item \textbf{Physiological evidence.} An analysis showing statistically significant differences in neurophysiological responses across sexism types and categories, highlighting aspects that remain challenging for content-only systems and providing insights into their current limitations.

    \item \textbf{A fusion model.} A hierarchical attention architecture that integrates physiological data with enriched content features, yielding significant, robust improvements over strong baselines on multiple sexism-detection tasks, with the largest benefits on nuanced, ambiguous cases.
\end{enumerate}

\vspace{-6mm}
\section{Related Work}

The automatic detection of harmful content is a well-established field of research. It is crucial to distinguish between \textbf{hate speech}, a broad concept \cite{nockleby2000}; \textbf{misogyny}, which involves hatred or dislike of women; and \textbf{sexism}, which refers to prejudice or discrimination based on sex and can manifest subtly. Misogyny lies at the intersection of hate speech and sexism, sharing characteristics of both.

Sexism detection has evolved significantly, moving from textual analysis to multimodal and multilingual approaches, driven by shared tasks such as \chg{task 10 at SemEval-2023} \cite{kirk2023semeval} and, more recently, the EXIST lab series \cite{plaza2025exist}, which has expanded its focus to include memes and videos. Methodologies have progressed through several stages:

\begin{enumerate}
    \item \textbf{Statistical Learning:} Early approaches used features like TF-IDF and N-grams with classical models such as SVM and Logistic Regression \cite{jha2017compliment, de2017offensive}.
    \item \textbf{Word Embedding-Based:} Later, neural networks (CNN, LSTM) fed with pre-trained embeddings (Word2Vec, GloVe) improved the capture of semantic context \cite{gasparini2018multimodal, grosz2020automatic}.
    \item \textbf{Pre-trained Language Models (PLMs):} The advent of Transformers (BERT, RoBERTa) marked a qualitative leap, enabling deep contextual understanding \cite{parikh2021categorizing}. Multimodal models (e.g., Visual-BERT) began to more effectively integrate textual and visual information \cite{rizzi2023recognizing, arcos2024sexism}.
    \item \textbf{Large Language Models (LLMs):} The most recent models, like ChatGPT and Llama, are used for their advanced reasoning capabilities, although their performance on specific classification tasks can vary compared to fine-tuned PLMs \cite{li2024hot, abercrombie2024revisiting}.
\end{enumerate}

Despite these advances, key challenges persist. Models often lack generalizability to unseen data \cite{samory2021call} and suffer from \textbf{biases} inherited from training data. Interpretability remains a problem, as most models function as "black boxes," making it difficult to understand their decisions.

In parallel, a line of research has emerged that integrates physiological data into Natural Language Processing (NLP) to capture human cognitive and affective responses.\chg{:}
\begin{itemize}
    \item \textbf{Eye-Tracking} has been used to understand cognitive load and attention, showing that gaze patterns shift when encountering complex, ambiguous, or emotionally salient content \cite{Khurana_2023, Das_2025}. 
    \item \textbf{Heart Rate} provides a general indicator of physiological arousal and stress. In our study, HR was continuously recorded to obtain descriptive measures of cardiac activity during meme-viewing, allowing temporal alignment with EEG and eye-tracking data \cite{Azarbarzin_2014}.
    \item \textbf{Electroencephalography} offers a direct, high-temporal-resolution measurement of brain activity, allowing for the identification of neural correlates of semantic and emotional processing in real time \cite{Lin_2024}.
\end{itemize}

While these sensors have been used in NLP, they have not addressed the specific ambiguity of sexist memes. We treat physiological responses as complementary content-level signals: eye-tracking, heart rate, and electroencephalography features are aggregated across viewers and aligned with the \emph{majority} EXIST labels per meme, in order to make the model learn generalized patterns linked to consensually sexist content. These signals do not replace human annotations; they help disambiguate cases where text–image cues are insufficient, in line with recent work aligning models with human perceptual and cognitive signals \cite{lrec2024_interead, lrec2020_zuco}. However, this approach also entails important limitations: models may encounter conflicting information when physiological reactions from a small number of subjects diverge from the majority labels, introducing potential noise and reducing the reliability of aggregated physiological patterns. For instance, \citet{lrec2024_interead} introduced \textit{InteRead}, a 50-subject eye-tracking dataset of real-world texts with systematically triggered interruptions, providing fine-grained annotations of gaze behavior and resumption lags. Their analyses confirmed that interruptions, as well as lexical variables such as word length and frequency, significantly shape reading dynamics, highlighting gaze as a robust indicator of cognitive load. Similarly, \citet{lrec2020_zuco} released \textit{ZuCo 2.0}, which records simultaneous EEG and eye-tracking during both natural reading and annotation tasks, thereby capturing neural and ocular correlates of comprehension and task-specific processing. Together, these resources demonstrate the feasibility and scientific value of integrating physiological signals into NLP pipelines, underscoring the timeliness of our proposal to incorporate ET, HR, and EEG into sexism detection models, moving beyond purely textual or visual features.

\vspace{-3mm}
\section{A Multimodal Resource of Sexist Meme Perception}
To build and test our human-centered models, we created a new multimodal resource by collecting physiological responses to existing, expertly annotated sexist and non-sexist memes.

\subsection{EXIST 2025 Memes}
We used the official training split of the EXIST 2025 shared task \cite{plaza2025exist}, a publicly available dataset containing \chg{3984} memes (\chg{1979} in Spanish, and \chg{2005} in English). In EXIST, sexist memes are further categorized by communicative intent: \textbf{Direct sexism} refers to content that itself expresses or promotes sexist ideas, whereas \textbf{Judgmental sexism} describes content that portrays or condemns sexist situations or behaviors. The memes were annotated in EXIST for three hierarchical tasks, with statistics summarized in Table \ref{tab:dataset_stats}.

Each meme was annotated by six annotators, revealing a clear trend in subjectivity. While \textbf{Task 1} (binary sexism detection) achieved fair agreement (Fleiss' $\kappa = 0.282$), the consensus dropped significantly for more nuanced judgments. For \textbf{Task 2} (source intention), agreement was merely slight ($\kappa \approx 0.19$). This subjectivity was most pronounced in \textbf{Task 3} (fine-grained categorization), where the \textit{Misogyny and Non-Sexual Violence} category showed the lowest agreement of all ($\kappa = 0.103$). This high degree of human disagreement on the content’s meaning strongly motivates our physiological approach, which seeks a more objective signal of perception. For our models, we used the majority-vote labels provided by the task organizers, and all experiments reported in this paper were conducted on the training partition.

\begin{table}[t!] \centering \scriptsize \caption{EXIST 2025 Meme Dataset Statistics (TRAIN). Percentages for T2 and T3 are calculated over the sexist subset of each language.} \label{tab:dataset_stats} \adjustbox{max width=\columnwidth}{ \begin{tabular}{llrr} \toprule \textbf{Task} & \textbf{Label} & \textbf{Spanish (N, \%)} & \textbf{English (N, \%)} \\ \midrule \multirow{3}{*}{T1: Sexism ID} & Sexist & \chg{1040 (52.6)} & \chg{963 (48.0)} \\ & Non-Sexist & \chg{626 (31.6)} & \chg{741 (37.0)} \\ & Ties & \chg{313 (15.8)} & \chg{301 (15.0)} \\ \midrule \multirow{3}{*}{T2: Intention} & Direct & \chg{731 (70.3)} & \chg{575 (59.7)} \\ & Judgmental & \chg{179 (17.2)} & \chg{226 (23.5)} \\ & Ties & \chg{130 (12.5)} & \chg{162 (16.8)} \\ \midrule \multirow{6}{*}{T3: Category} & Ideological Ineq. & \chg{290 (27.9)} & \chg{296 (30.7)} \\ & Stereotyping & \chg{249 (23.9)} & \chg{202 (21.0)} \\ & Objectification & \chg{228 (21.9)} & \chg{212 (22.0)} \\ & Sexual Violence & \chg{77 (7.4)} & \chg{60 (6.2)} \\ & Misogyny NSV & \chg{32 (3.1)} & \chg{19 (2.0)} \\ & Ties & \chg{164 (15.8)} & \chg{174 (18.1)} \\ \midrule \multicolumn{2}{l}{\textbf{Total Memes}} & \textbf{\chg{1979}} & \textbf{\chg{2005}} \\ \bottomrule \end{tabular} } \end{table}

\vspace{-3mm}
\subsection{Physiological Data Collection}
We conducted two experiments with a total of 16 subjects of different nationalities, with a gender-balanced distribution\chg{, in their 20s and 30s. This age range was considered particularly relevant, as younger populations have been reported to exhibit greater tolerance toward sexist content~\cite{plaza2024exist}, enabling the investigation of their unconscious cognitive and emotional reactions to such content}. Each of the \chg{3984} memes was viewed by at least two subjects.

\vspace{-3mm}
\paragraph{Procedure.} Subjects were seated comfortably while memes were displayed on a screen until they provided a response, followed by a 3-second pause before the next stimulus to prevent overlap. In each session ($\approx$ 45 minutes), subjects viewed ~100-170 memes. After each meme, they answered control questions about its content to ensure engagement and comprehension. All subjects gave their consent to use the data anonymously for research purposes.

\chg{We used the following devices to collect physiological data during the experiments:}

\begin{itemize}
    \item \textbf{Eye-Tracking:} Pupil Labs Neon glasses recorded binocular gaze data at 200 Hz.
    \item \textbf{Heart-Rate:} A Garmin Venu 3 watch continuously measured inter-beat intervals.
    \item \textbf{Electroencephalography:} A 16-channel OpenBCI Cyton Ultracortex Mark IV headset recorded neural activity from 16 scalp locations (10–20 system) at 250 Hz.
\end{itemize}

\subsection{Physiological Feature Extraction}
From the raw physiological data, we extracted a comprehensive feature set for each meme-viewing instance, structured by experimental setup.

\paragraph{\textbf{Experiment 1 (ET)}}
\begin{itemize}
    \item \textbf{ET:} For each oculomotor event (fixations, blinks, pupil diameter), we computed summary statistics (mean, standard deviation, minimum, maximum, count) capturing detailed patterns of visual attention.
\end{itemize}

\paragraph{\textbf{Experiment 2 (EEG)}}
\begin{itemize}
    \item \textbf{EEG:} Signals were acquired concurrently with HR. Preprocessing included: (1) conversion to $\mu$V; (2) zero-phase 4th-order Butterworth band-pass filter (0.5–40 Hz); and (3) baseline correction using a 2-second pre-stimulus interval. For each of the 16 channels (10–20 system), we extracted time-domain statistics and frequency-domain features by estimating PSDs with Welch’s method (256-sample windows) and integrating power over Delta (0.5–4 Hz), Theta (4–8 Hz), Alpha (8–13 Hz), Beta (13–30 Hz), and Gamma (30–40 Hz). Features were harmonized via Box–Cox transformation, ComBat correction \cite{FORTIN2018104}, Winsorization, and robust Z-scoring.
\end{itemize}

\paragraph{\textbf{Common measures (HR and RT)}}
\begin{itemize}
    \item \textbf{HR:} Recorded in both experiments to provide a shared physiological reference and enable temporal alignment across modalities. For each trial, we computed summary statistics (mean, standard deviation, minimum, maximum) as indicators of overall arousal.

   \item \textbf{Reaction Time (RT):} Recorded in both experiments, defined as the elapsed time between the initial presentation of a meme and the subject’s response to proceed to the next stimulus. Stimuli remained on screen until response, followed by a 3-second pause to prevent overlap.
\end{itemize}

This process yielded a rich, multimodal dataset linking meme content to synchronized physiological responses, forming our new resource.

\vspace{-3mm}
\section{Physiological Correlates of Sexism Perception}
Our analysis (\textbf{RQ1}) revealed significant physiological markers, showing that the human brain and body appear to respond with specific patterns to sexist content.
\vspace{-2mm}
\subsection{Cognitive Load and Autonomic Arousal}
Eye-tracking data established a clear cognitive load gradient. As shown in Table \ref{tab:physio_results_expanded}, both \textbf{reaction time} and \textbf{fixation count} increased significantly and progressively ($p < .001$) from non-sexist memes, to direct sexist memes, and finally to judgmental sexist memes, which required the most cognitive effort. Blink duration was also significantly shorter for direct sexist content ($p=.005$), consistent with heightened visual attention, as shorter blink durations have been shown to occur under increased visual workload and sustained attentional engagement \cite{Benedetto2011}.

\begin{table*}[t!]
\centering
\small
\caption{Key physiological metrics across sexism levels (Mean ± SD). $p$-values from ANOVA.}
\label{tab:physio_results_expanded}
\begin{tabular}{lcccc}
\toprule
\textbf{Metric} & \textbf{Non-Sexist} & \textbf{Direct Sexist} & \textbf{Judgmental} & \textbf{p-value} \\
\midrule
Reaction Time (s) & 13.68$\pm$9.10 & 15.84$\pm$11.40 & 17.58$\pm$12.08 & \pval{0.000} \\
Fixation Count & 40.31$\pm$28.37 & 44.42$\pm$33.67 & 50.34$\pm$37.61 & \pval{0.000} \\
Blink Duration (ms) & 267.36$\pm$57.65 & 261.13$\pm$55.27 & 263.05$\pm$50.58 & \pval{0.007} \\
\bottomrule
\end{tabular}
\end{table*}
\vspace{-2mm}
\subsection{Category-Specific and Neural Signatures}
A granular analysis revealed unique neurophysiological signatures for specific types of sexism. Most notably, memes featuring \textbf{Objectification} triggered a significant \textbf{constriction of the right pupil} ($p=.036$), a known physiological marker of processing aversive visual stimuli, as pupillary constriction has been shown to occur in response to unpleasant or emotionally negative content independent of luminance or arousal effects \cite{Ayzenberg_2018, Blini_2023}.

\begin{figure*}[t!]
    \centering
    \begin{subfigure}[b]{0.48\linewidth}
        \includegraphics[width=\linewidth]{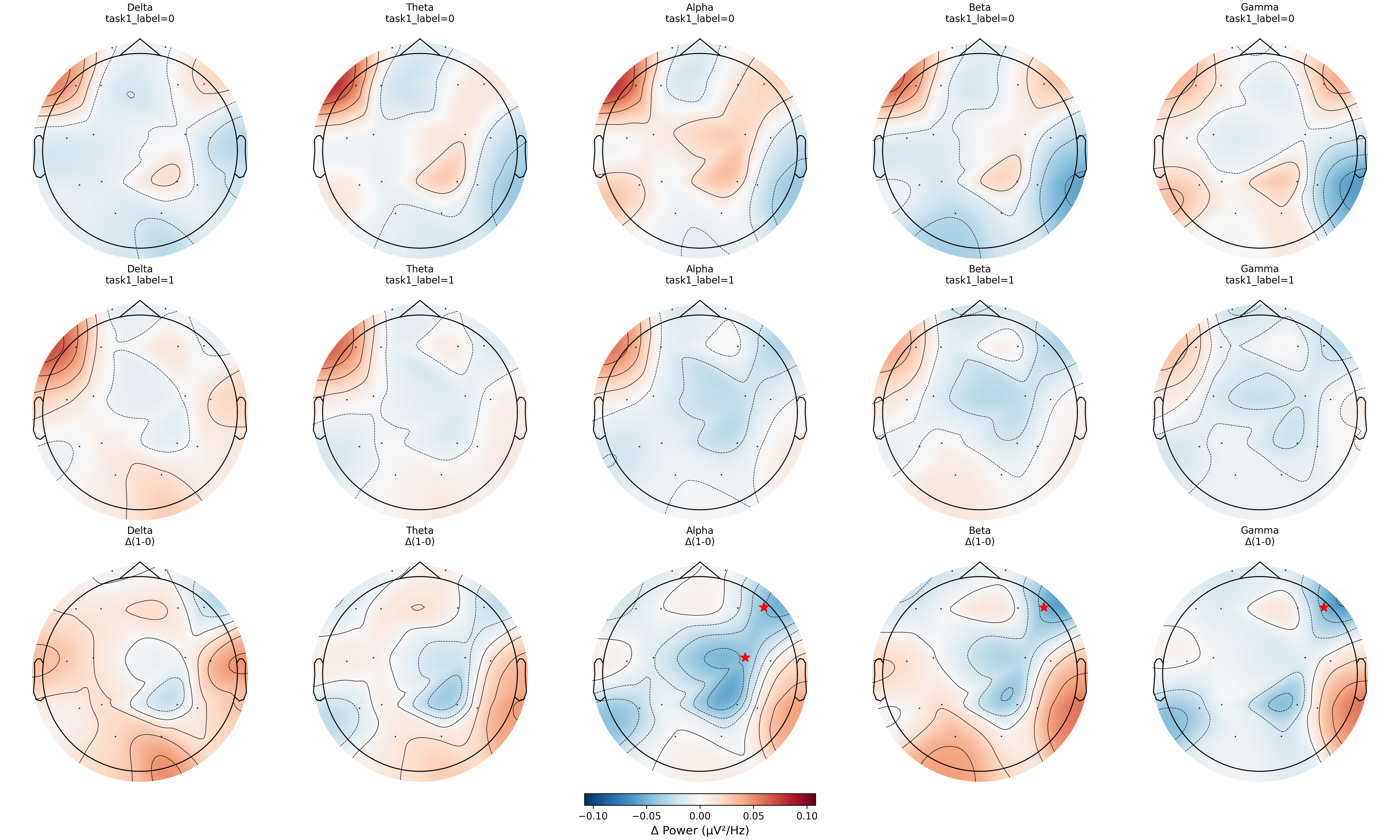}
        \caption{Task 1: Sexist vs. Non-Sexist}
        \label{fig:topo_task1}
    \end{subfigure}
    \hfill
    \begin{subfigure}[b]{0.48\linewidth}
        \includegraphics[width=\linewidth]{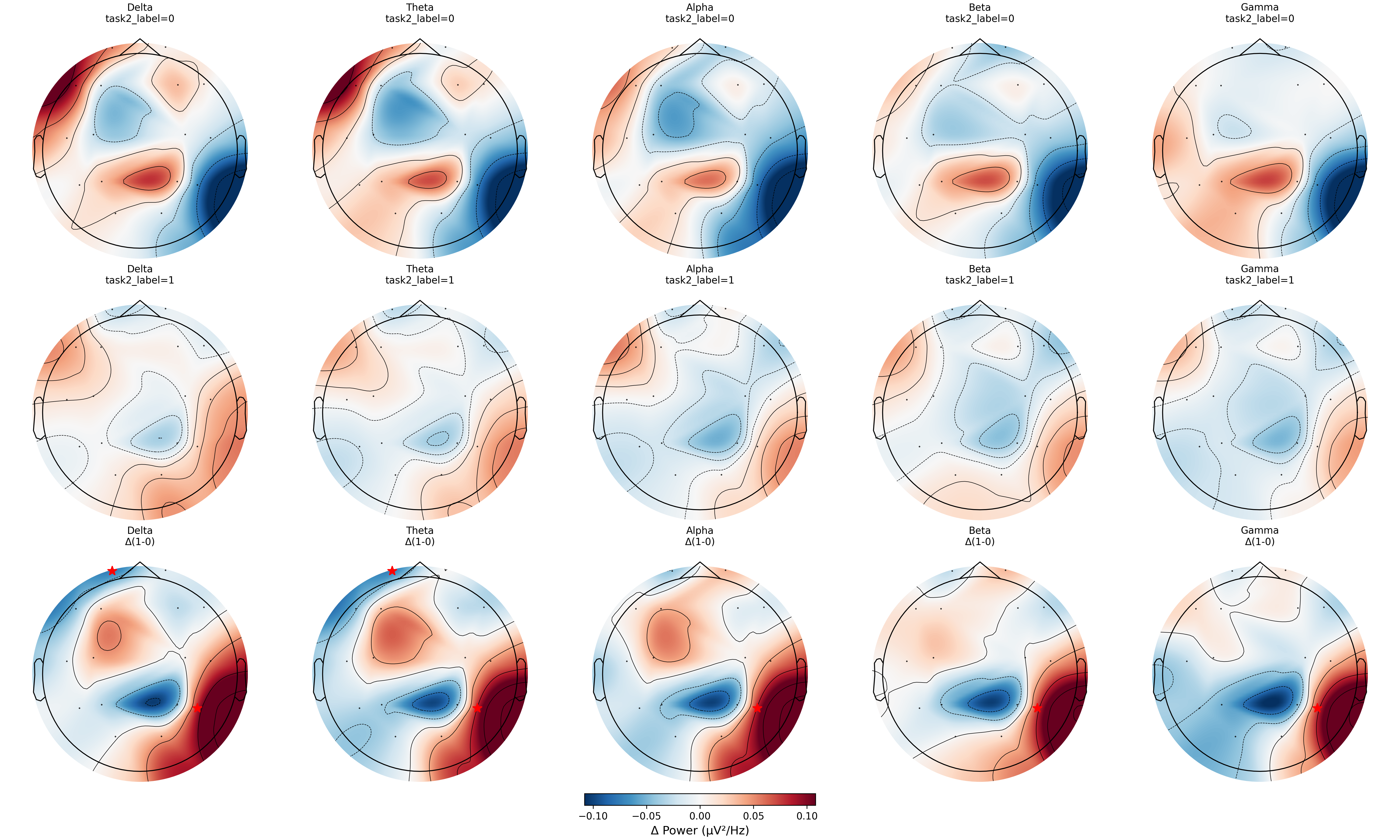}
        \caption{Task 2: Direct vs. Judgmental}
        \label{fig:topo_task2}
    \end{subfigure}

    \vspace{1em}

    \begin{subfigure}[b]{0.48\linewidth}
        \includegraphics[width=\linewidth]{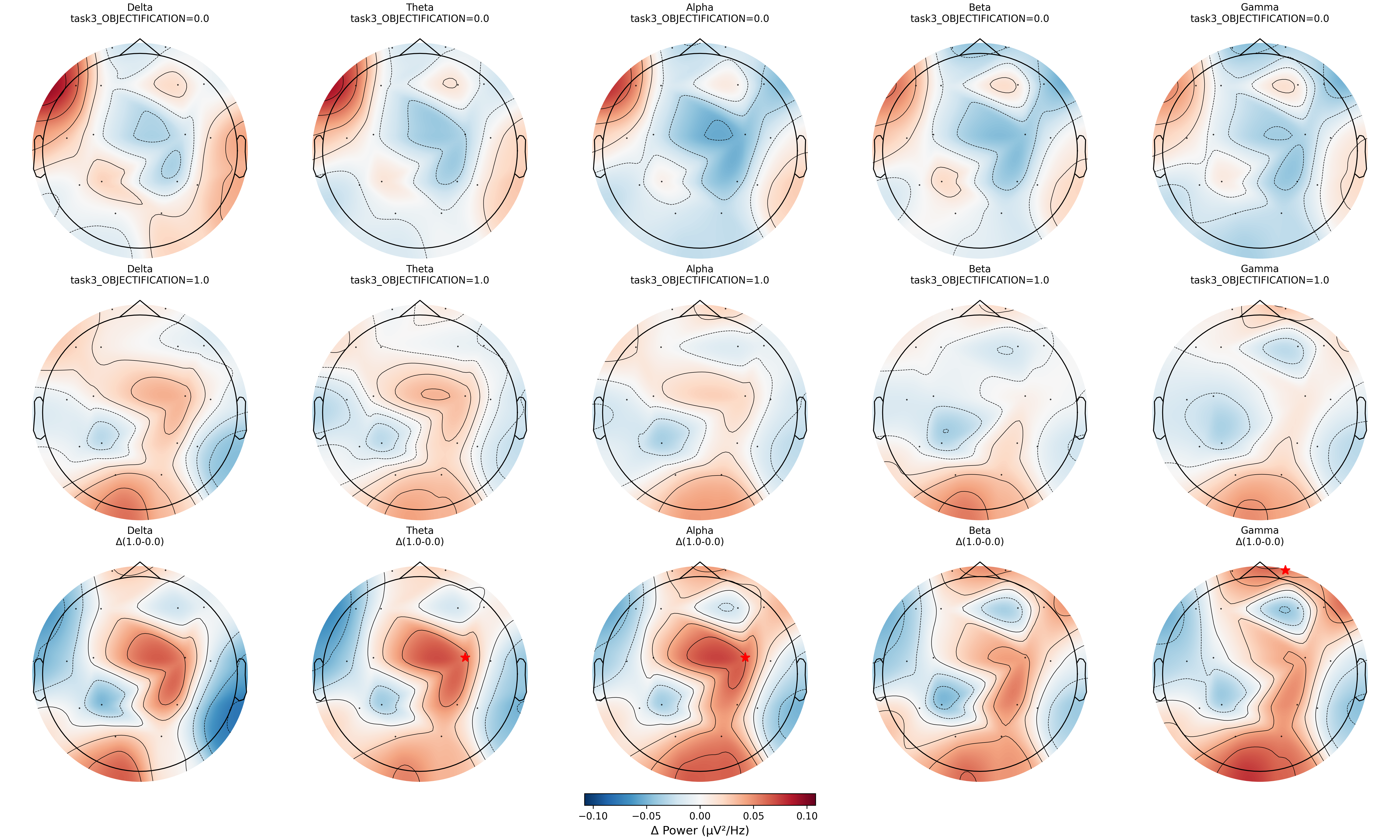}
        \caption{Task 3: Objectification}
        \label{fig:topo_objectification}
    \end{subfigure}
    \hfill
    \begin{subfigure}[b]{0.48\linewidth}
        \includegraphics[width=\linewidth]{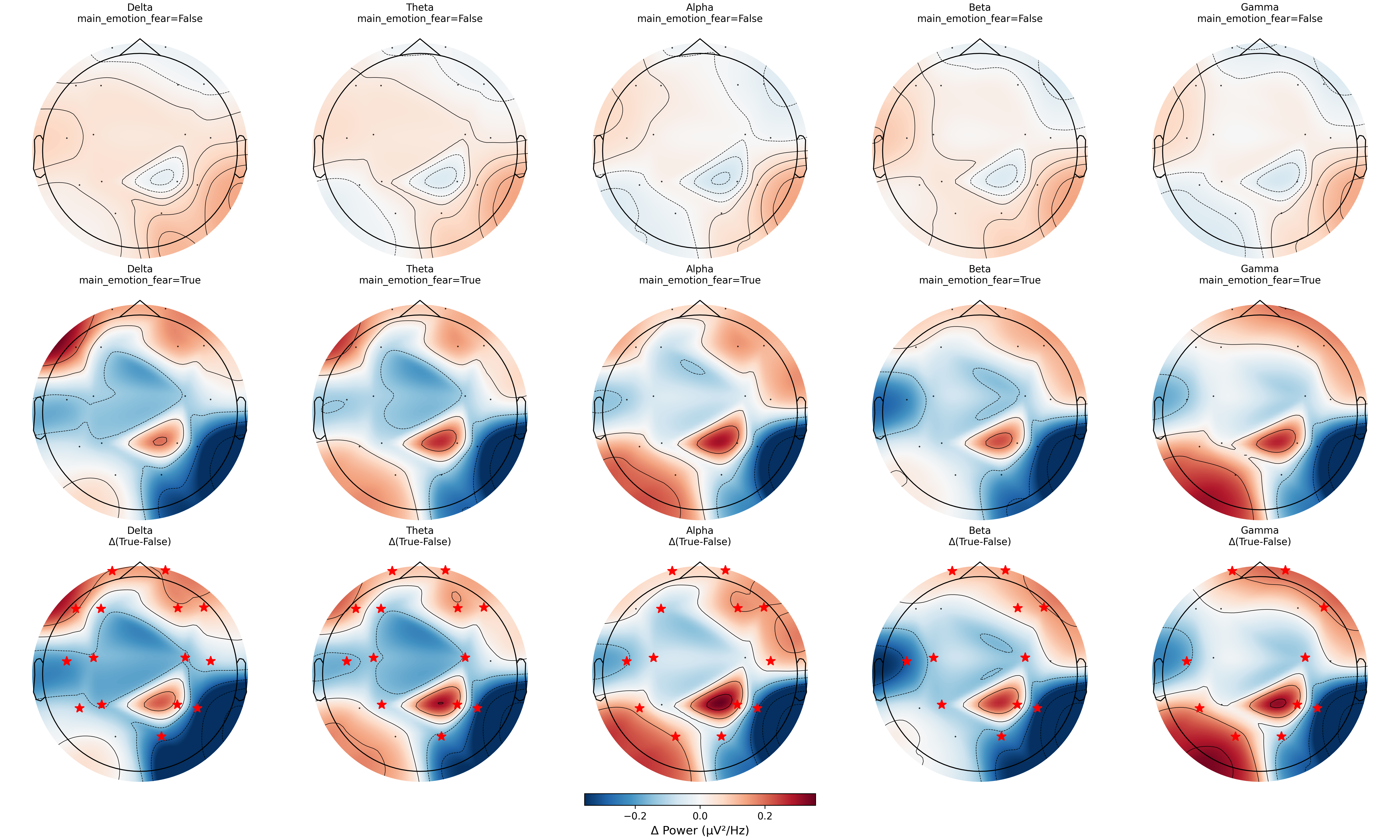}
        \caption{Emotion: Fear}
        \label{fig:topo_fear}
    \end{subfigure}
    \caption{EEG topographic maps of band-power differences across key experimental contrasts. 
For each subfigure, the \textbf{top row} shows the mean power for the first condition and the \textbf{middle row} for the second; the \textbf{bottom row} shows their difference (\textit{Condition 2 – Condition 1}). Specifically: 
(a) Non-Sexist vs. Sexist; 
(b) Judgmental vs. Direct Sexism; 
(c) Non-Objectification vs. Objectification; 
(d) Neutral vs. Fear (OCR-based emotion). 
Columns correspond to Delta, Theta, Alpha, Beta, and Gamma bands. 
Red indicates power increase, blue indicates power decrease; red stars mark channels with statistically significant differences ($p<0.05$).}

    \label{fig:topoplots_combined}
\end{figure*}

EEG analysis offered insights into brain activity. For binary sexism detection (\textbf{Task 1}), we observed a significant \textbf{power reduction (desynchronization) in the Alpha, Beta, and Gamma bands} over right frontal (anterior) channels when viewing sexist compared to non-sexist memes. This pattern is consistent with prior work on affective picture processing reporting alpha–beta desynchronization and gamma-band modulation in both posterior and anterior (including frontal) regions \cite{Schubring_2019,Strube_2021}. The right-frontal localization in our data may reflect higher-order evaluative or conflict-monitoring processes engaged by sexist content. (Figure \ref{fig:topoplots_combined}a). When distinguishing \textit{direct} from \textit{judgmental} sexism (\textbf{Task 2}), viewing direct memes triggered a \textbf{widespread power increase (synchronization) across multiple frequency bands} in the right parietal region, consistent with prior evidence showing enhanced gamma-band synchronization in parietal and visual cortices during the processing of emotionally salient or hostile visual stimuli \cite{Luo_2009,Headley_2013}, suggesting more intense sensory and affective processing for overtly hostile content (Figure \ref{fig:topoplots_combined}b). The analysis of fine-grained categories and emotions revealed further unique signatures. For instance, memes categorized as \textbf{Objectification} were associated with significant power increases in Alpha/Theta at central channel C4 and Gamma at frontal Fp2 (Figure \ref{fig:topoplots_combined}c). Finally, when the meme's OCR text was automatically classified as inducing \textbf{Fear}\footnote{Emotion labels were derived from the meme's OCR text using language-specific models: \texttt{daveni/twitter-xlm-roberta-emotion-es} (Spanish) and \texttt{j-hartmann/emotion-english-distilroberta-base} (English).}, the EEG registered widespread, bilateral frontal activation across all bands, a pattern consistent with prior findings linking frontal–parietal oscillatory dynamics to threat detection and heightened vigilance during emotionally salient or fear-inducing contexts \cite{Grimshaw_2014,Bodala_2016} (Figure \ref{fig:topoplots_combined}d).

\begin{figure*}[t!]
    \centering
    \includegraphics[width=0.95\textwidth]{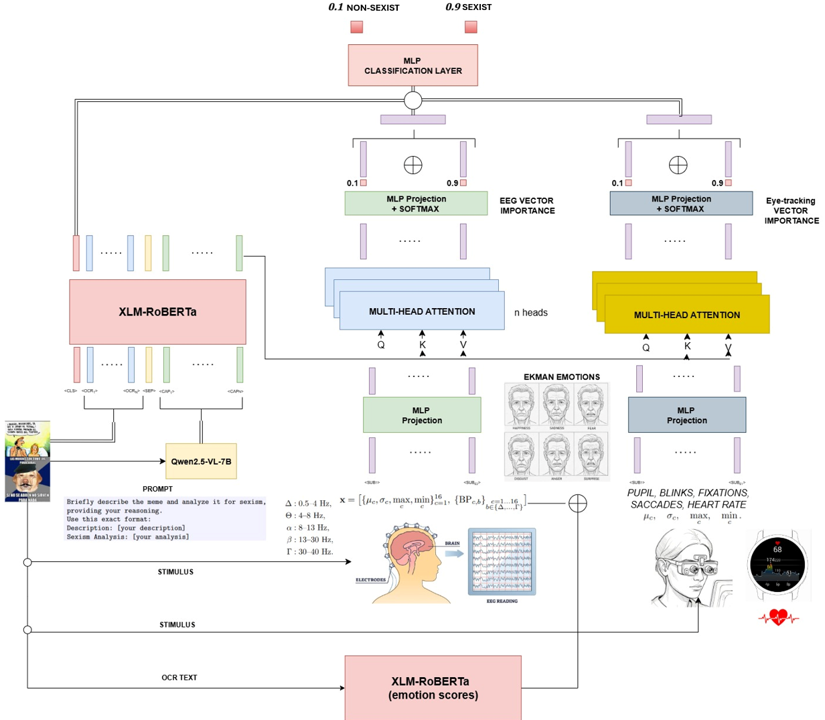}
    \caption{Architecture of the final hierarchical attention-based fusion model. It integrates enriched text (OCR + VLM caption) with sequences of physiological reactions from several subjects (EEG and Eye-Tracking/HR). Cross-attention mechanisms allow the model to learn correlations between specific textual tokens and physiological responses.}
    \label{fig:model_architecture}
\end{figure*}
\vspace{-2mm}
\section{Multimodal Fusion Model}

To leverage these rich signals, we developed a hierarchical attention-based fusion model (Figure \ref{fig:model_architecture}) designed to integrate enriched content features with sequences of physiological responses.
\vspace{-2mm}
\paragraph{Enriched Content Representation.} Instead of directly fusing raw image features with text, which can be ineffective, we first use a Vision-Language Model (\texttt{Qwen2.5-VL-7B}\footnote{https://huggingface.co/Qwen/Qwen2.5-VL-7B-Instruct}) to generate a detailed textual description and preliminary sexism analysis of each meme. This generated text is concatenated with the meme's original OCR text, separated by a special token. 

\paragraph{Physiological Fusion Core.} The EEG and ET/HR feature vectors are treated as sequences (where each token represents one subject's reaction). Each physiological sequence is independently passed through a multi-head cross-attention layer. In this crucial step, the physiological sequence acts as the \textbf{Query}, while the text token embeddings from the XLM-RoBERTa (XLM-R) encoder serve as the \textbf{Key} and \textbf{Value}. This allows the model to learn which textual elements are most correlated with the observed neural and attentional reactions. The resulting text-aware physiological representations are aggregated via a learned weighted sum and concatenated with the main text embedding (\texttt{[CLS]} token) for final classification through a small Multilayer Perceptron (MLP) head.
\vspace{-3mm}
\paragraph{Training Strategy.} The model is trained in two phases for stability: first, only the fusion head and attention layers are trained for 5 epochs with the XLM-R backbone frozen (LR = 5e-5). Then, the entire model is fine-tuned for 10 epochs using discriminative learning rates (2e-6 for lower layers, 1e-5 for upper layers, 5e-5 for the head) with the AdamW optimizer and a weighted Binary Cross-Entropy (BCE) loss function to handle class imbalance.
\vspace{-3mm}
\section{Results and Analysis}

We evaluated our models using 5-fold cross-validation, reporting macro F1-score and AUC. Our unimodal analysis (\textbf{RQ2}) showed that EEG-based classifiers performed surprisingly well, achieving an AUC of 0.717 for Task 1, on par with the text-only XLM-R (AUC 0.704). This confirms that physiological data alone is a powerful predictor of perceived sexism.

\begin{figure*}[t!]
    \centering
    \includegraphics[width=0.7\textwidth]{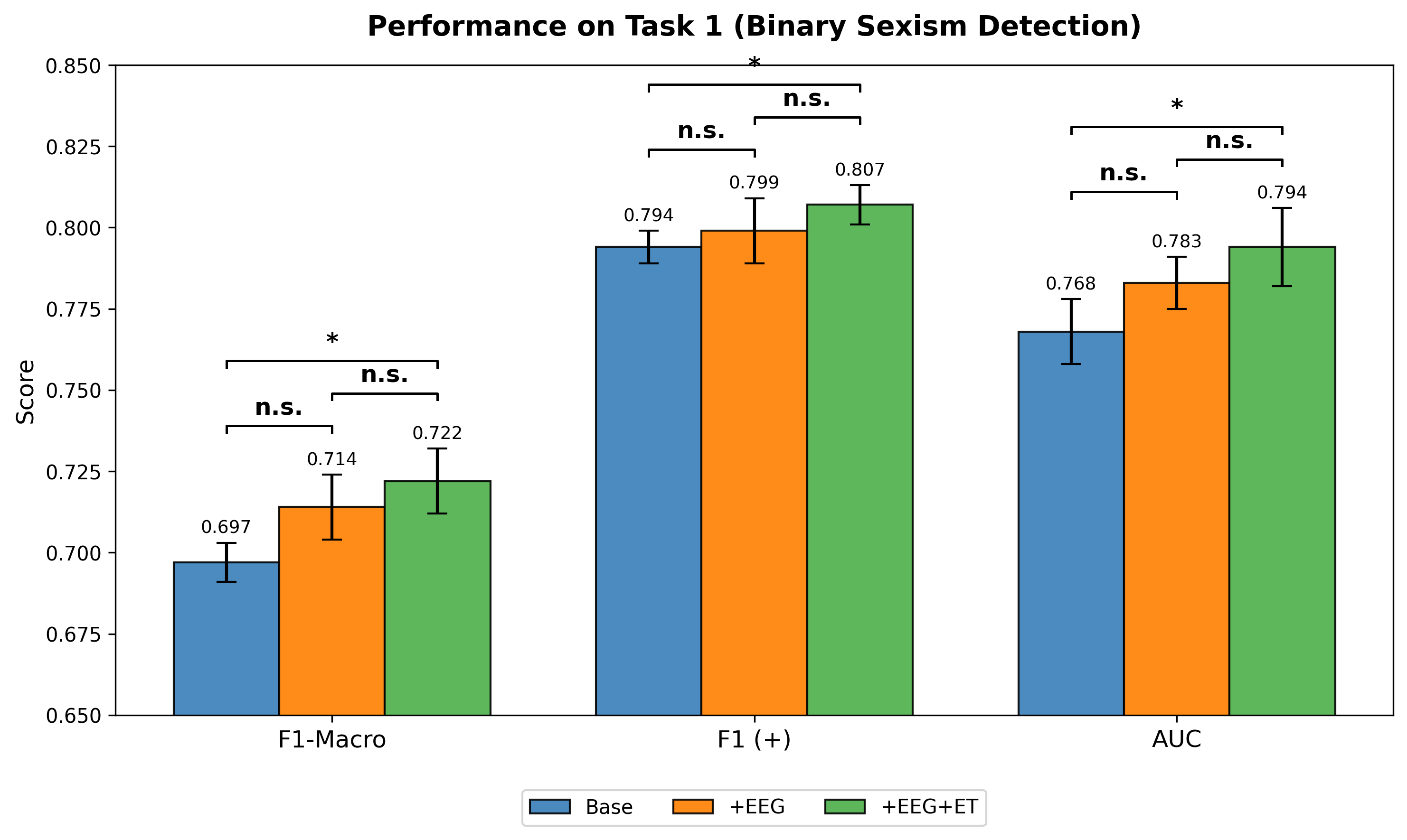}
    \caption{Performance on Task 1 (Binary Sexism Detection) with 95\% confidence intervals. Bars represent model performance scores (Macro F1, F1+, and AUC), showing a progressive and statistically significant improvement ($p < 0.05$) as all physiological signals are added.}
    \label{fig:significance_plot}
\end{figure*}

Our multimodal results (\textbf{RQ3}), summarized in Table \ref{tab:multimodal_results}, demonstrate the synergistic power of our fusion strategy. Our content-only baseline (\texttt{Qwen-VL + XLM-R}) is already very strong, significantly outperforming a simple ViT+XLM-R concatenation. However, incrementally adding physiological signals yields statistically significant improvements across all tasks, confirmed by non-overlapping 95\% confidence intervals (Figure \ref{fig:significance_plot}).

For binary sexism detection (\textbf{Task 1}), our full model achieves an AUC of \textbf{0.794}, a 3.4\% relative improvement over the strong baseline. For the highly nuanced intention detection (\textbf{Task 2}), where human agreement is low, the physiological signals provide a crucial disambiguating signal, boosting AUC by 4.3\% from 0.628 to \textbf{0.655}.

\begin{table}[t!]
\centering
\scriptsize
\setlength{\tabcolsep}{4pt}
\caption{Multimodal model performance (AUC) across all tasks. Improvements from the baseline are statistically significant ($p<.05$).}
\label{tab:multimodal_results}
\begin{tabularx}{\columnwidth}{@{}Xccc@{}}
\toprule
\textbf{Model / Fusion} & \textbf{T1: Binary} & \textbf{T2: Intention} & \textbf{T3: Category} \\
\midrule
(Simple Fusion) & & & \\
XLM-R + ViT & $0.699 \pm .012$ & $0.575 \pm .028$ & $0.703 \pm .031$ \\
\midrule
(Our Models) & & & \\
Qwen-VL + XLM-R & $0.768 \pm .016$ & $0.628 \pm .017$ & $0.782 \pm .023$ \\
+ EEG & $0.783 \pm .013$ & $0.634 \pm .019$ & $0.779 \pm .020$ \\
+ EEG + ET/HR & $\mathbf{0.794 \pm .019}$ & $\mathbf{0.655 \pm .025}$ & $\mathbf{0.784 \pm .015}$ \\
\bottomrule
\end{tabularx}
\end{table}

The most dramatic impact is in fine-grained categorization (\textbf{Task 3}). As shown in Table \ref{tab:task3_f1_results}, the F1-score for the most challenging class, \textit{Misogyny \& Non-Sexual Violence}, jumps by an unprecedented \textbf{26.3\%} (from 0.259 to 0.327) with the full model. This shows that when content features are weak, the objective signal of the human response becomes indispensable for more accurate classification.

\begin{table}[t!]
\centering
\scriptsize
\caption{F1-score for Task 3 categories. The full model significantly improves classification accuracy on the hardest classes.}
\label{tab:task3_f1_results}
\begin{tabular}{lcc}
\toprule
\textbf{Category (Task 3)} & \textbf{Baseline F1} & \textbf{Full Model F1} \\
\midrule
Ideological Inequality & $0.566 \pm .020$ & $\mathbf{0.585 \pm .007}$ \\
Stereotyping Dominance & $0.479 \pm .051$ & $\mathbf{0.498 \pm .020}$ \\
Objectification & $0.541 \pm .037$ & $\mathbf{0.561 \pm .026}$ \\
Sexual Violence & $0.396 \pm .045$ & $\mathbf{0.428 \pm .041}$ \\
Misogyny NSV & $0.259 \pm .074$ & $\mathbf{0.327 \pm .033}$ \\
\midrule
\textbf{Macro Average} & $0.448 \pm .045$ & $\mathbf{0.480 \pm .025}$ \\
\bottomrule
\end{tabular}
\end{table}

\subsection{Attention-Based Interpretability}

\begin{figure}[t!]
    \centering

    \begin{subfigure}[b]{0.8\linewidth}
        \centering
        \includegraphics[width=0.6\linewidth]{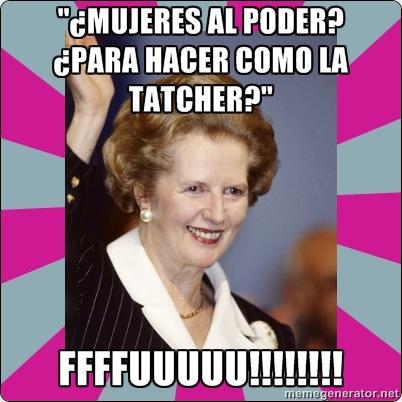}
        \caption{Original meme (ID 110051).}
        \label{fig:attention_meme}
    \end{subfigure}

    \vspace{0.5em}

    \begin{subfigure}[b]{0.9\linewidth}
        \centering
        \includegraphics[width=0.7\linewidth]{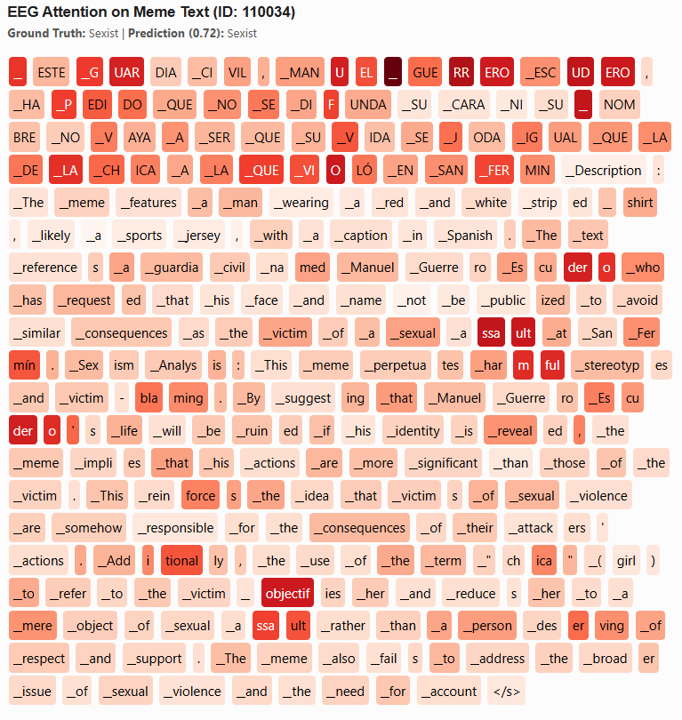}
        \caption{Cross-modal attention map between EEG signals and textual tokens.}
        \label{fig:attention_map}
    \end{subfigure}

    \caption{Attention-based interpretation of a correctly classified sexist meme.}
    \label{fig:attention_example_single}
\end{figure}

To explore how the multimodal fusion model integrates physiological and textual information, we visualized its cross-attention mechanisms. Figure \ref{fig:attention_example_single} presents a case study of a subtle sexist meme that the content-only baseline misclassified but our full model correctly identified as sexist. The meme (Figure \ref{fig:attention_example_single}a) shows a well-known female political leader with the Spanish caption “¿MUJERES AL PODER? ¿PARA HACER COMO LA TATCHER? FFFFUUUU!!!!” (“WOMEN IN POWER? TO DO LIKE THATCHER? FFFFUUUU!!!!”). The meme uses sarcasm, taking an extreme example of female leadership to ridicule the idea of women in power. It expresses frustration toward women’s political participation, reflecting ideological or stereotypical rather than overt misogynistic sexism.

The attention map (Figure \ref{fig:attention_example_single}b) illustrates how the model integrates multimodal cues. High attention weights align with the meme's literal text, particularly “¿MUJERES AL PODER?”, and with analytical phrases generated by the VLM, such as “critique of women's ability to lead.” These associations suggest that the model learns to connect implicit neural patterns with explicit semantic representations of sexism. Note that this visualization does not represent a direct physiological mapping between EEG activity and specific words. It reflects correlations learned by the model, not temporally aligned neural responses. This should be viewed as an AI-level inference rather than a physiological interpretation, which we acknowledge as a limitation that future work should address by integrating EEG with ET for word-level alignment.

\vspace{-2mm}

\section{Conclusion and Future Work}
This work validates a novel, human-centered paradigm for sexism detection. The results seem to indicate that physiological responses to memes provide a rich, objective signal of perception that complements content analysis. We have collected and will release a first-of-its-kind multimodal resource pairing memes with ET, HR, and EEG data, opening new avenues for research on affective computing, focused on systems that sense and interpret human \chg{cognitive and emotional reactions}. Our multimodal fusion model, which learns to associate implicit physiological patterns with explicit content features, delivers consistent gains over strong content-only baselines, indicating that aligning multimodal systems with human \chg{psychological} responses is an effective strategy for building more accurate and robust classifiers. Notably, no significant effects were observed for heart rate, likely because static and brief meme stimuli may not elicit sufficient autonomic activation \cite{Brouwer_2013}. A further limitation is that memes were not controlled for visual or textual complexity, which could also influence physiological responses independently of sexist content.

Future work should focus on modeling the rich subjectivity in our data, moving beyond majority-vote labels to predict user-specific or demographic-aware perceptions of harm using frameworks like Learning with Disagreements \cite{uma-etal-2021-semeval}. Extending this paradigm to dynamic video content (e.g., from TikTok), which introduces temporal complexities, is another critical next step. By continuing to "listen" to the nuanced signals of the human brain and body, we can build more accurate, robust, and ultimately more human-aware systems to foster safer digital environments.

\section{Ethical Considerations}

The physiological data was gathered from 16 subjects of different nationalities that gave their consent to use the data anonymously for research purposes. The meme stimuli, while sometimes offensive, were part of a pre-existing dataset used for the EXIST shared tasks. For the analysis, the demographic and physiological data were anonymized to protect subject privacy. The goal of this research is to develop \chg{human-centered AI models}, and we advocate for the use of the dataset and models introduced in this work in systems that prioritize transparency, user agency, and robust appeals processes.


\begin{acknowledgments}
This work was done in the framework of the research project \textit{Multimodal and Multisensor data-centric AI models (ANNOTATE-MULTI2)} funded by MICIU/AEI/10.13039/501100011033 and by ERDF/EU (Grant PID2024-156022OB-C32). The work of Elena~Gomis-Vicent was carried out while she was at the PRHLT Research Center.
\end{acknowledgments}

\section{References}
\label{sec:reference}

\bibliographystyle{lrec2026-natbib}
\bibliography{sample-ceur}

@String{Computing = "Computing" }

@String{Macmillan = "Macmillan" }

@String{Springer = "Springer-Verlag" }

@InProceedings{plaza2025exist,
  author    = {Plaza, Laura and Carrillo-de-Albornoz, Jorge and Arcos, Iv{\'a}n and Rosso, Paolo and Spina, Damiano and Amig{\'o}, Enrique and Gonzalo, Julio and Morante, Roser},
  editor    = {Hauff, Claudia and Macdonald, Craig and Jannach, Dietmar and Kazai, Gabriella and Nardini, Franco Maria and Pinelli, Fabio and Silvestri, Fabrizio and Tonellotto, Nicola},
  title     = "{{EXIST} 2025: Learning with disagreement for sexism identification and characterization in tweets, memes, and {TikTok} videos}",
  booktitle = {Advances in Information Retrieval},
  year      = {2025},
  publisher = {Springer Nature Switzerland},
  address   = {Cham},
  pages     = {442--449},
  isbn      = {978-3-031-88720-8},
  doi       = {10.1007/978-3-031-88720-8_65}
}

@misc{geh_ucsd,
  author       = {Dehingia, Nabamallika},
  title        = {When social media is sexist: A call to action against online gender-based violence},
  howpublished = {\url{https://geh.ucsd.edu/when-social-media-is-sexist-a-call-to-action-against-online-gender-based-violence/}},
  month        = may,
  day          = 13,
  year         = {2020},
}

@misc{nomore_org,
  author       = {Swan, Teia},
  title        = {Online Misogyny: A Digital Threat},
  howpublished = {The KNOW MORE Blog, NO MORE Project},
  year         = {2025},
  month        = feb,
  day          = {11},
  note         = {Accessed: 11 Sept. 2025},
  url          = {https://www.nomore.org/online-misogyny-a-digital-threat/}
}

@techreport{unwomen_repository,
  author       = {{United Nations Entity for Gender Equality and the Empowerment of Women (UN Women)}},
  title        = {Repository of {UN} Women’s Work on Technology-Facilitated Violence Against Women and Girls},
  type         = {Catalogue/Directory},
  institution  = {Ending Violence Against Women Section, UN Women},
  publisher    = {United Nations Entity for Gender Equality and the Empowerment of Women (UN Women)},
  month        = mar,
  year         = {2025},
  pages        = {12},
  url          = {https://www.unwomen.org/sites/default/files/2025-03/repository-of-un-womens-work-on-technology-facilitated-violence-against-women-and-girls-en.pdf},
  note         = {Accessed: July 11, 2025}
}

@misc{amnesty_uk_toxic_tech,
  author       = {{Amnesty International UK}},
  title        = {Toxic tech: New polling exposes widespread online misogyny driving Gen Z away from social media},
  howpublished = {Press release, Amnesty International UK},
  year         = {2025},
  month        = mar,
  day          = {21},
  url          = {https://www.amnesty.org.uk/press-releases/toxic-tech-new-polling-exposes-widespread-online-misogyny-driving-gen-z-away-social}
}

@misc{issac_karmaveer_sexism_memes,
  author       = {Issac, Aalen Chacko},
  title        = {Reinforcement of Sexism Through Memes},
  howpublished = {Report at National Dialogue on Gender-based Cyber Violence},
  year         = {2018},
  note         = {Retrieved from IT for Change website: \url{https://projects.itforchange.net/e-vaw/wp-content/uploads/2018/01/Aalen_Issac-2.pdf}}
}

@book{nockleby2000,
  title={Hate Speech},
  author={John T. Nockleby},
  booktitle={Encyclopedia of the American Constitution},
  edition={2},
  editor={Leonard W. Levy, Kenneth L. Karst, et al.},
  pages={1277--1279},
  year={2000},
  publisher={Macmillan},
  address={New York}
}

@article{reunir_sexist_humour,
  author  = {Argüello-Gutiérrez, Catalina and Cubero, Ana and Fumero, Fabiola and Montealegre, Diana and Sandoval, Pía and Smith-Castro, Vanessa},
  year    = {2022},
  month   = {10},
  pages   = {},
  title   = {I'm just joking! {Perceptions} of sexist humor and sexist beliefs in a Latin American Context},
  volume  = {Online first},
  journal = {International Journal of Psychology},
  doi     = {10.1002/ijop.12884}
}

@inproceedings{ut_cs_female_astronaut_misogyny,
  title = "{{F}emale Astronaut: Because sandwiches won{'}t make themselves up there}: Towards multimodal misogyny detection in memes",
  author = {Singh, Smriti and Haridasan, Amritha and Mooney, Raymond},
  editor = {Chung, Yi-ling and Röttger, Paul and Nozza, Debora and Talat, Zeerak and Mostafazadeh Davani, Aida},
  booktitle = {The 7th Workshop on Online Abuse and Harms (WOAH)},
  month = jul,
  year = {2023},
  address = {Toronto, Canada},
  publisher = {Association for Computational Linguistics},
  url = {https://aclanthology.org/2023.woah-1.15/},
  doi = {10.18653/v1/2023.woah-1.15},
  pages = {150--159}
}

@InProceedings{kiela2020hateful,
  author    = {Kiela, Douwe and Firooz, Hamed and Mohan, Aravind and Goswami, Vedanuj and Singh, Amanpreet and Ringshia, Pratik and Testuggine, Davide},
  title     = "{The {Hateful Memes} Challenge: Detecting hate speech in multimodal memes}",
  booktitle = {Proceedings of the 34th International Conference on Neural Information Processing Systems},
  year      = {2020},
  publisher = {Curran Associates Inc.},
  address   = {Red Hook, NY, USA},
  location  = {Vancouver, BC, Canada},
  series    = {{NeurIPS} '20},
  articleno = {220},
  numpages  = {14},
  isbn      = {9781713829546},
  doi       = {10.48550/arXiv.2005.04790},
  url       = {https://doi.org/10.48550/arXiv.2005.04790}
}

@inproceedings{arxiv_explaining_sexism_detection,
    title = "Explaining Matters: Leveraging Definitions and Semantic Expansion for Sexism Detection",
    author = "Khan, Sahrish  and
      Jhumka, Arshad  and
      Pergola, Gabriele",
    editor = "Che, Wanxiang  and
      Nabende, Joyce  and
      Shutova, Ekaterina  and
      Pilehvar, Mohammad Taher",
    booktitle = "Proceedings of the 63rd Annual Meeting of the Association for Computational Linguistics (Volume 1: Long Papers)",
    month = jul,
    year = "2025",
    address = "Vienna, Austria",
    publisher = "Association for Computational Linguistics",
    url = "https://aclanthology.org/2025.acl-long.809/",
    doi = "10.18653/v1/2025.acl-long.809",
    pages = "16553--16571",
    ISBN = "979-8-89176-251-0",
}

@article{acl_anthology_gaze4hate,
  author    = {Bradley, Margaret M. and Miccoli, Laura and Escrig, Miguel A. and Lang, Peter J.},
  title     = {The pupil as a measure of emotional arousal and autonomic activation},
  journal   = {Psychophysiology},
  year      = {2008},
  volume    = {45},
  number    = {4},
  pages     = {602--607},
  doi       = {10.1111/j.1469-8986.2008.00654.x},
  pmid      = {18282202},
  pmcid     = {PMC3612940}
}

@article{plosone_rhrveasy,
author = {García Martínez, Constantino and Bardají, Sofía and Pérez-Tirador, Pablo and Otero, Abraham},
year = {2024},
month = {11},
pages = {},
title = {RHRVEasy: Heart rate variability made easy},
volume = {19},
journal = {PLOS ONE},
doi = {10.1371/journal.pone.0309055}
}

@article{researchgate_physiological_signals_affective_computing,
author = {Quan, X.-L and Zeng, Z.-G and Jiang, J.-H and Zhang, Y.-Q and Lv, B.-L and Wu, D.-R},
year = {2021},
month = {08},
pages = {1769-1784},
title = {Physiological Signals Based Affective Computing: A Systematic Review},
volume = {47},
journal = {Zidonghua Xuebao/Acta Automatica Sinica},
doi = {10.16383/j.aas.c200783}
}

@inproceedings{lrec2020_zuco,
    title = "{Z}u{C}o 2.0: A Dataset of Physiological Recordings During Natural Reading and Annotation",
    author = "Hollenstein, Nora  and
      Troendle, Marius  and
      Zhang, Ce  and
      Langer, Nicolas",
    editor = "Calzolari, Nicoletta  and
      B{\'e}chet, Fr{\'e}d{\'e}ric  and
      Blache, Philippe  and
      Choukri, Khalid  and
      Cieri, Christopher  and
      Declerck, Thierry  and
      Goggi, Sara  and
      Isahara, Hitoshi  and
      Maegaard, Bente  and
      Mariani, Joseph  and
      Mazo, H{\'e}l{\`e}ne  and
      Moreno, Asuncion  and
      Odijk, Jan  and
      Piperidis, Stelios",
    booktitle = "Proceedings of the Twelfth Language Resources and Evaluation Conference",
    month = may,
    year = "2020",
    address = "Marseille, France",
    publisher = "European Language Resources Association",
    url = "https://aclanthology.org/2020.lrec-1.18/",
    pages = "138--146",
    language = "eng",
    ISBN = "979-10-95546-34-4",
}

@article{Azarbarzin_2014,
author = {Azarbarzin, Ali and Ostrowski, Michele and Hanly, Patrick and Younes, Magdy},
year = {2014},
month = {06},
pages = {645-653},
title = {Relationship between Arousal Intensity and Heart Rate Response to Arousal},
volume = {37},
journal = {Sleep},
doi = {10.5665/sleep.3560}
}

@article{Ayzenberg_2018,
  author    = {Ayzenberg, Vladislav and Hickey, Meghan R. and Lourenco, Stella F.},
  title     = {Pupillometry reveals the physiological underpinnings of the aversion to holes},
  journal   = {PeerJ},
  year      = {2018},
  volume    = {6},
  pages     = {e4185},
  doi       = {10.7717/peerj.4185},
  pmid      = {29312818},
  pmcid     = {PMC5756615},
  url       = {https://doi.org/10.7717/peerj.4185}
}

@article{Headley_2013,
  author  = {Headley, Drew B. and Paré, Denis},
  title   = {In sync: gamma oscillations and emotional memory},
  journal = {Frontiers in Behavioral Neuroscience},
  year    = {2013},
  volume  = {7},
  pages   = {170},
  doi     = {10.3389/fnbeh.2013.00170},
  pmid    = {24319416},
  pmcid   = {PMC3836200},
  url     = {https://doi.org/10.3389/fnbeh.2013.00170}
}

@article{Luo_2009,
  author  = {Luo, Qian and Mitchell, Derek and Cheng, Xi and Mondillo, Krystal and McCaffrey, Daniel and Holroyd, Tom and Carver, Frederick and Coppola, Richard and Blair, James},
  title   = {Visual awareness, emotion, and gamma band synchronization},
  journal = {Cerebral Cortex},
  year    = {2009},
  volume  = {19},
  number  = {8},
  pages   = {1896--1904},
  doi     = {10.1093/cercor/bhn216},
  pmid    = {19047574},
  pmcid   = {PMC2705698},
  url     = {https://doi.org/10.1093/cercor/bhn216}
}

@article{Grimshaw_2014,
  author  = {Grimshaw, Gina M. and Foster, Joshua J. and Corballis, Paul M.},
  title   = {Frontal and parietal EEG asymmetries interact to predict attentional bias to threat},
  journal = {Brain and Cognition},
  year    = {2014},
  volume  = {90},
  pages   = {76--86},
  doi     = {10.1016/j.bandc.2014.06.008},
  pmid    = {25014408}
}

@article{Bodala_2016,
  author  = {Bodala, Indu P. and Li, Junhua and Thakor, Nitish V. and Al-Nashash, Hasan},
  title   = {EEG and Eye Tracking Demonstrate Vigilance Enhancement with Challenge Integration},
  journal = {Frontiers in Human Neuroscience},
  year    = {2016},
  volume  = {10},
  pages   = {273},
  doi     = {10.3389/fnhum.2016.00273},
  pmid    = {27378868},
  pmcid   = {PMC4914593}
}

@article{Schubring_2019,
  author    = {Schubring, David and Schupp, Harald T.},
  title     = {Affective picture processing: Alpha- and lower beta-band desynchronization reflects emotional arousal},
  journal   = {Psychophysiology},
  year      = {2019},
  volume    = {56},
  number    = {8},
  pages     = {e13386},
  doi       = {10.1111/psyp.13386},
  pmid      = {31026079},
  url       = {https://doi.org/10.1111/psyp.13386}
}

@article{Brouwer_2013,
  author    = {Anne-Marie Brouwer and Nelleke van Wouwe and Christian M{\"u}hl and Jan B. F. van Erp and Alexander Toet},
  title     = {Perceiving blocks of emotional pictures and sounds: effects on physiological variables},
  journal   = {Frontiers in Human Neuroscience},
  year      = {2013},
  volume    = {7},
  pages     = {295},
  doi       = {10.3389/fnhum.2013.00295},
  pmid      = {23801957},
  pmcid     = {PMC3689025},
  url       = {https://doi.org/10.3389/fnhum.2013.00295}
}

@article{Benedetto2011,
title = {Driver workload and eye blink duration},
journal = {Transportation Research Part F: Traffic Psychology and Behaviour},
volume = {14},
number = {3},
pages = {199-208},
year = {2011},
issn = {1369-8478},
doi = {https://doi.org/10.1016/j.trf.2010.12.001},
url = {https://www.sciencedirect.com/science/article/pii/S136984781000094X},
author = {Simone Benedetto and Marco Pedrotti and Luca Minin and Thierry Baccino and Alessandra Re and Roberto Montanari}}

@article{Strube_2021,
  author    = {Strube, Andreas and Rose, Michael and Fazeli, Sepideh and B{\"u}chel, Christian},
  title     = {Alpha-to-beta- and gamma-band activity reflect predictive coding in affective visual processing},
  journal   = {Scientific Reports},
  year      = {2021},
  volume    = {11},
  pages     = {24059},
  doi       = {10.1038/s41598-021-02939-z},
  pmid      = {34873255},
  pmcid     = {PMC8648824},
  url       = {https://doi.org/10.1038/s41598-021-02939-z}
}

@article{Blini_2023,
  author    = {Blini, Elvio and Zorzi, Marco},
  title     = {Pupil size as a robust marker of attentional bias toward nicotine-related stimuli in smokers},
  journal   = {Psychonomic Bulletin \& Review},
  year      = {2023},
  volume    = {30},
  number    = {2},
  pages     = {596--607},
  doi       = {10.3758/s13423-022-02192-z},
  url       = {https://doi.org/10.3758/s13423-022-02192-z}
}

@inproceedings{lrec2024_interead,
    title = "{I}nte{R}ead: An Eye Tracking Dataset of Interrupted Reading",
    author = {Zermiani, Francesca  and
      Dhar, Prajit  and
      Sood, Ekta  and
      K{\"o}gel, Fabian  and
      Bulling, Andreas  and
      Wirzberger, Maria},
    editor = "Calzolari, Nicoletta  and
      Kan, Min-Yen  and
      Hoste, Veronique  and
      Lenci, Alessandro  and
      Sakti, Sakriani  and
      Xue, Nianwen",
    booktitle = "Proceedings of the 2024 Joint International Conference on Computational Linguistics, Language Resources and Evaluation (LREC-COLING 2024)",
    month = may,
    year = "2024",
    address = "Torino, Italia",
    publisher = "ELRA and ICCL",
    url = "https://aclanthology.org/2024.lrec-main.802/",
    pages = "9154--9169",
}

@InProceedings{kirk2023semeval,
  author    = {Kirk, Hannah and Yin, Wenjie and Vidgen, Bertie and R{\"o}ttger, Paul},
  editor    = {Ojha, Atul Kr. and Do{\u{g}}ru{\"o}z, A. Seza and Da San Martino, Giovanni and Tayyar Madabushi, Harish and Kumar, Ritesh and Sartori, Elisa},
  title     = "{S}em{E}val-2023 Task 10: Explainable Detection of Online Sexism",
  booktitle = {Proceedings of the 17th International Workshop on Semantic Evaluation ({SemEval}-2023)},
  month     = jul,
  year      = {2023},
  address   = {Toronto, Canada},
  publisher = {Association for Computational Linguistics},
  pages     = {2193--2210},
  doi       = {10.18653/v1/2023.semeval-1.305},
  url       = {https://aclanthology.org/2023.semeval-1.305}
}

@inproceedings{de2017offensive,
  author    = {Pelle, Rogers Prates and Moreira, Viviane P.},
  title     = {Offensive Comments in the Brazilian Web: a Dataset and Baseline Results},
  booktitle = {Proceedings of the 6th Brazilian Workshop on Social Network Analysis and Mining (BraSNAM)},
  year      = {2017},
  address   = {São Paulo, Brazil},
  pages     = {510--519},
  publisher = {Sociedade Brasileira de Computação},
  doi       = {10.5753/brasnam.2017.3260},
  issn      = {2595-6094}
}

@InProceedings{gasparini2018multimodal,
  author    = {Gasparini, Francesca and Fersini, Elisabetta and Perj{\'e}si, Zsolt and Anzovino, Marianna and Messina, Emanuela and Pievani, Manuela},
  title     = "{Multimodal classification of sexist advertisements}",
  booktitle = {Proceedings of the 15th International Joint Conference on e-Business and Telecommunications (ICETE)},
  volume    = {1},
  pages     = {399--406},
  year      = {2018},
  organization = {SciTePress},
  doi       = {10.5220/0006859403990406},
  url       = {https://www.scitepress.org/papers/2018/68594/68594.pdf}
}

@inproceedings{grosz2020automatic,
author = {Grosz, Dylan and Conde-Cespedes, Patricia},
title = {Automatic Detection of Sexist Statements Commonly Used at the Workplace},
year = {2020},
isbn = {978-3-030-60469-1},
publisher = {Springer-Verlag},
address = {Berlin, Heidelberg},
url = {https://doi.org/10.1007/978-3-030-60470-7_11},
doi = {10.1007/978-3-030-60470-7_11},
booktitle = {Trends and Applications in Knowledge Discovery and Data Mining: PAKDD 2020 Workshops, DSFN, GII, BDM, LDRC and LBD, Singapore, May 11–14, 2020, Revised Selected Papers},
pages = {104–115},
numpages = {12},
location = {Singapore, Singapore}
}

@article{samory2021call,
  title={{C}all me sexist, but... : Revisiting Sexism Detection Using Psychological Scales and Adversarial Samples},
  volume={15},
  url={https://ojs.aaai.org/index.php/ICWSM/article/view/18085},
  DOI={10.1609/icwsm.v15i1.18085},
  number={1},
  journal={Proceedings of the International AAAI Conference on Web and Social Media},
  author={Samory, Mattia and Sen, Indira and Kohne, Julian and Flöck, Fabian and Wagner, Claudia},
  year={2021},
  month={May},
  pages={573--584}
}

@inproceedings{jha2017compliment,
    title = "When does a compliment become sexist? Analysis and classification of ambivalent sexism using twitter data",
    author = "Jha, Akshita  and
      Mamidi, Radhika",
    editor = {Hovy, Dirk  and
      Volkova, Svitlana  and
      Bamman, David  and
      Jurgens, David  and
      O{'}Connor, Brendan  and
      Tsur, Oren  and
      Do{\u{g}}ru{\"o}z, A. Seza},
    booktitle = "Proceedings of the Second Workshop on {NLP} and Computational Social Science",
    month = aug,
    year = "2017",
    address = "Vancouver, Canada",
    publisher = "Association for Computational Linguistics",
    url = "https://aclanthology.org/W17-2902/",
    doi = "10.18653/v1/W17-2902",
    pages = "7--16",
}

@article{FORTIN2018104,
title = {Harmonization of cortical thickness measurements across scanners and sites},
journal = {NeuroImage},
volume = {167},
pages = {104-120},
year = {2018},
issn = {1053-8119},
doi = {https://doi.org/10.1016/j.neuroimage.2017.11.024},
url = {https://www.sciencedirect.com/science/article/pii/S105381191730931X},
author = {Jean-Philippe Fortin and Nicholas Cullen and Yvette I. Sheline and Warren D. Taylor and Irem Aselcioglu and Philip A. Cook and Phil Adams and Crystal Cooper and Maurizio Fava and Patrick J. McGrath and Melvin McInnis and Mary L. Phillips and Madhukar H. Trivedi and Myrna M. Weissman and Russell T. Shinohara},
}

@InProceedings{arcos2024sexism,
  author    = {Arcos, Iv{\'a}n and Rosso, Paolo},
  editor    = {Goeuriot, Lorraine and Mulhem, Philippe and Qu{\'e}not, Georges and Schwab, Didier and Di Nunzio, Giorgio Maria and Soulier, Laure and Galu{\v{s}}{\v{c}}{\'a}kov{\'a}, Petra and Garc{\'i}a Seco de Herrera, Alba and Faggioli, Guglielmo and Ferro, Nicola},
  title = "{Sexism identification on {TikTok}: A multimodal {AI} approach with text, audio, and video}",
  booktitle = {Experimental IR Meets Multilinguality, Multimodality, and Interaction},
  year      = {2024},
  publisher = {Springer Nature Switzerland},
  address   = {Cham},
  pages     = {61--73},
  isbn      = {978-3-031-71736-9},
  doi       = {10.1007/978-3-031-71736-9_2}
}

@InProceedings{plaza2024exist,
author="Plaza, Laura
and Carrillo-de-Albornoz, Jorge
and Ruiz, V{\'i}ctor
and Maeso, Alba
and Chulvi, Berta
and Rosso, Paolo
and Amig{\'o}, Enrique
and Gonzalo, Julio
and Morante, Roser
and Spina, Damiano",
editor="Goeuriot, Lorraine
and Mulhem, Philippe
and Qu{\'e}not, Georges
and Schwab, Didier
and Di Nunzio, Giorgio Maria
and Soulier, Laure
and Galu{\v{s}}{\v{c}}{\'a}kov{\'a}, Petra
and Garc{\'i}a Seco de Herrera, Alba
and Faggioli, Guglielmo
and Ferro, Nicola",
title="Overview of EXIST 2024 --- Learning with Disagreement for Sexism Identification and Characterization in Tweets and Memes",
booktitle="Experimental IR Meets Multilinguality, Multimodality, and Interaction",
year="2024",
publisher="Springer Nature Switzerland",
address="Cham",
pages="93--117",
isbn="978-3-031-71908-0"
}

@article{parikh2021categorizing,
author = {Parikh, Pulkit and Abburi, Harika and Chhaya, Niyati and Gupta, Manish and Varma, Vasudeva},
title = {Categorizing Sexism and Misogyny through Neural Approaches},
year = {2021},
issue_date = {November 2021},
publisher = {Association for Computing Machinery},
address = {New York, NY, USA},
volume = {15},
number = {4},
issn = {1559-1131},
url = {https://doi.org/10.1145/3457189},
doi = {10.1145/3457189},
journal = {ACM Trans. Web},
month = jun,
articleno = {17},
numpages = {31}
}

@article{rizzi2023recognizing,
title = {Recognizing misogynous memes: Biased models and tricky archetypes},
journal = {Information Processing \& Management},
volume = {60},
number = {5},
pages = {103474},
year = {2023},
issn = {0306-4573},
doi = {https://doi.org/10.1016/j.ipm.2023.103474},
url = {https://www.sciencedirect.com/science/article/pii/S030645732300211X},
author = {Giulia Rizzi and Francesca Gasparini and Aurora Saibene and Paolo Rosso and Elisabetta Fersini},
}

@article{li2024hot,
author = {Li, Lingyao and Fan, Lizhou and Atreja, Shubham and Hemphill, Libby},
 title = {{"HOT" {ChatGPT}: The promise of {ChatGPT} in detecting and discriminating hateful, offensive, and toxic comments on social media}},
year = {2024},
issue_date = {May 2024},
publisher = {Association for Computing Machinery},
address = {New York, NY, USA},
volume = {18},
number = {2},
issn = {1559-1131},
url = {https://doi.org/10.1145/3643829},
doi = {10.1145/3643829},
journal = {ACM Trans. Web},
month = mar,
articleno = {30},
numpages = {36}
}

@inproceedings{abercrombie2024revisiting,
    title = "Revisiting Annotation of Online Gender-Based Violence",
    author = "Abercrombie, Gavin  and
      Vitsakis, Nikolas  and
      Jiang, Aiqi  and
      Konstas, Ioannis",
    editor = "Abercrombie, Gavin  and
      Basile, Valerio  and
      Bernadi, Davide  and
      Dudy, Shiran  and
      Frenda, Simona  and
      Havens, Lucy  and
      Tonelli, Sara",
    booktitle = "Proceedings of the 3rd Workshop on Perspectivist Approaches to NLP (NLPerspectives) @ LREC-COLING 2024",
    month = may,
    year = "2024",
    address = "Torino, Italia",
    publisher = "ELRA and ICCL",
    url = "https://aclanthology.org/2024.nlperspectives-1.3/",
    pages = "31--41",
}

@inproceedings{Khurana_2023,
    title = "Synthesizing Human Gaze Feedback for Improved {NLP} Performance",
    author = "Khurana, Varun  and
      Kumar, Yaman  and
      Hollenstein, Nora  and
      Kumar, Rajesh  and
      Krishnamurthy, Balaji",
    editor = "Vlachos, Andreas  and
      Augenstein, Isabelle",
    booktitle = "Proceedings of the 17th Conference of the European Chapter of the Association for Computational Linguistics",
    month = may,
    year = "2023",
    address = "Dubrovnik, Croatia",
    publisher = "Association for Computational Linguistics",
    url = "https://aclanthology.org/2023.eacl-main.139/",
    doi = "10.18653/v1/2023.eacl-main.139",
    pages = "1895--1908"}

@inproceedings{Lin_2024,
author = {Peiran Lin and Liwen Yang},
title = {{Building a virtual psychological counselor by integrating EEG emotion detection with large-scale NLP models}},
volume = {12924},
booktitle = {Third International Conference on Biological Engineering and Medical Science (ICBioMed2023)},
editor = {Alan Wang},
organization = {International Society for Optics and Photonics},
publisher = {SPIE},
pages = {129241X},
year = {2024},
doi = {10.1117/12.3013169},
URL = {https://doi.org/10.1117/12.3013169}
}

@inproceedings{uma-etal-2021-semeval,
    title = "{S}em{E}val-2021 Task 12: Learning with Disagreements",
    author = "Uma, Alexandra  and
      Fornaciari, Tommaso  and
      Dumitrache, Anca  and
      Miller, Tristan  and
      Chamberlain, Jon  and
      Plank, Barbara  and
      Simpson, Edwin  and
      Poesio, Massimo",
    editor = "Palmer, Alexis  and
      Schneider, Nathan  and
      Schluter, Natalie  and
      Emerson, Guy  and
      Herbelot, Aurelie  and
      Zhu, Xiaodan",
    booktitle = "Proceedings of the 15th International Workshop on Semantic Evaluation (SemEval-2021)",
    month = aug,
    year = "2021",
    address = "Online",
    publisher = "Association for Computational Linguistics",
    url = "https://aclanthology.org/2021.semeval-1.41/",
    doi = "10.18653/v1/2021.semeval-1.41",
    pages = "338--347",
}

@article{Das_2025,
  author = {Kar, Avishake and Pathak, Ipsita and Mukherjee, Shibani and Chatterjee, Siddhartha},
  year = {2025},
  month = {07},
  pages = {423--428},
  title = {Decoding complexity and emotion: A computational linguistic approach to sentence comprehensibility and writer affect},
  isbn = {2456-2165},
  journal = {International Journal of Innovative Science and Research Technology},
  doi = {10.38124/ijisrt/25jul551}
}

\bibliographystylelanguageresource{lrec2026-natbib}
\bibliographylanguageresource{your_lr_bib_file_name}

\end{document}